\documentclass[12pt]{article} 

\usepackage{latexsym} % Gets \Box etc
\usepackage{amssymb}  % \gtrsim, \geqslant, etc etc: 
\usepackage{epsfig}       % For PostScript figures
\usepackage{multirow}  % for multirows in tables
\newcommand{\al}{\alpha}
\newcommand{\be}{\beta}
\newcommand{\ga}{\gamma}

\newcommand{\de}{\delta}
\newcommand{\De}{\Delta}
\newcommand{\ep}{\varepsilon}

\newcommand{\ka}{\kappa}

\newcommand{\La}{\Lambda}

\newcommand{\Si}{\Sigma}
\newcommand{\<}{\langle} 
\renewcommand{\>}{\rangle} % LaTeX: \> already defined

\newcommand{\txt}{\textstyle}

\newcommand\eqn[1]{(\ref{#1})}      % parentheses around the LaTex "ref" macro
\newcommand{\beq}{\begin{equation}}
\newcommand{\eeq}{\end{equation}}
\newcommand{\ba}{\begin{array}}
\newcommand{\bea}{\begin{eqnarray}}
\newcommand{\ea}{\end{array}}
\newcommand{\eea}{\end{eqnarray}}

\newcommand\comment[1]{ \hbox{[{\it Comment suppressed here.}\/]} }
\newcommand\hide[1]{}

\newcommand{\MeV}{{\rm MeV}}

\newcommand{\skipover}[1]{}

\newcommand{\half} {{\txt {1\over 2}}}

\makeatletter %\catcode`\@=11

\def\appendix{\par                              % Have \appendix say
    \setcounter{section}{0}                     % `Appendix A', not just `A'
    \setcounter{subsection}{0}
    \renewcommand{\theequation}{\Alph{section}.\arabic{equation}}
    \renewcommand{\thesection}{Appendix \Alph{section}
                \setcounter{equation}{0}  } %Have eqns numbered (A.1) etc
}

\def\applabel#1{\@bsphack
  \protected@write\@auxout{}%
         {\string\newlabel{#1}{{\Alph{section}}{\thepage}}}%
  \@esphack}
\def\section{
\setcounter{equation}{0}        % Reset eqn numbers at start of section
\@startsection {section}{1}{\z@}{-3.5ex plus -1ex minus 
 -.2ex}{2.3ex plus .2ex}{\large\bf}}
\renewcommand{\theequation}{\arabic{section}.\arabic{equation}}

\def\subsection{\@startsection{subsection}{2}{\z@}{-3.25ex plus -1ex minus 
 -.2ex}{1.5ex plus .2ex}{\normalsize\bf}}

\def\subsubsection{\@startsection{subsubsection}{3}{\z@}{-3.25ex plus
 -1ex minus -.2ex}{1.5ex plus .2ex}{\normalsize}}

\makeatother   %\catcode`\@=12

\newsavebox{\eqlabel}
\makeatletter  %\catcode`\@=11
\newlength{\numblen}
\newsavebox{\eqnumb}
\def\@eqnnum{\savebox{\eqnumb}{\rm (\theequation)}%
\settowidth{\numblen}{\usebox{\eqnumb}}%
\makebox[\numblen][l]{\usebox{\eqnumb}~~~\usebox{\eqlabel}}}
\makeatother   %\catcode`\@=12

\newenvironment{equationwithlabel}[1]{ %
  \begin{equation}\label{#1} }{\end{equation}} %\savebox{\eqlabel}{~}}
\newcommand{\beql}[1]{\begin{equationwithlabel}{#1}}
\newcommand{\eeql}{\end{equationwithlabel}}
\newcommand{\bt}{\begin{tabular}}
\newcommand{\et}{\end{tabular}}

\newcommand{\ms}{m_s}
\newcommand{\Ms}{M_s}

\newcommand{\Qt}{{\tilde Q}}
\newcommand{\Y}{{T_8}}

\begin{document}

\title{
{\bf Color superconductivity in dense quark matter}}

\author{
	Mark Alford \\[0.5ex]
{\normalsize Center for Theoretical Physics}\\
{\normalsize Massachusetts Institute of Technology}\\
%{\normalsize 77 Mass Ave }\\ 
{\normalsize Cambridge, MA 02139 }\\
{\tt\normalsize alford@mit.edu }
}

\newcommand{\preprintno}{
  \normalsize MIT-CTP-2962 % \\ DRAFT VERSION
}

\date{\today \\[1ex] \preprintno}

\begin{titlepage}
\maketitle
\def\thepage{}          % No page number on title page

\begin{abstract}
%       1         2         3         4         5         6
I discuss recent developments in our understanding of the
color-superconducting phases of cold, dense quark matter.  I describe
the phase diagram as a function of density and the strange quark mass,
and outline some ideas about possible observational consequences of
these exotic phases.
\end{abstract}

\bigskip\noindent
Contribution to the proceedings of the TMU-Yale symposium on the
dynamics of gauge fields, Tokyo metropolitan university, Dec 1999.

\end{titlepage}

\renewcommand{\thepage}{\arabic{page}}

\section{Introduction}
\label{sec:int}

The phase diagram of QCD is a topic of intensive research, both
theoretical and experimental. In recent years, the high-temperature
low-density regime has been studied in lattice gauge calculations, and
probed in heavy-ion experiments~\cite{mga_berndt}.  The low-temperature
high-density regime has remained more mysterious, because lattice
calculations run into intractable problems at non-zero chemical
potential, and heavy-ion collision experiments are focussing on high
temperature in order to definitively produce the quark-gluon plasma.
Even so, cold dense quark matter is of direct physical relevance (in
neutron stars, for example), and recently there has been striking
theoretical progress in our understanding of its symmetries and basic
properties.  In this paper I will review some of the elements of the
new picture that is emerging.

At high densities and low temperatures, matter consists of a Fermi sea
of quarks. The relevant degrees of freedom are those which involve
quarks with momenta near the Fermi surface. These interact via
gluons, in a manner described by QCD. The quark-quark interaction
has two channels available, the antisymmetric $\bar{\bf 3}$,
and the symmetric {\bf 6}. It is attractive in the $\bar{\bf 3}_A$:
this can be seen from single-gluon-exchange, or
by counting of strings, or from the 'tHooft vertex
induced by instantons.

It was shown by Bardeen, Cooper, and Schrieffer (BCS)~\cite{mga_BCS} that a
Fermi surface in the presence of attractive interactions is unstable.
If there is {\em any} channel in which the quark-quark interaction is
attractive, then the true ground state of the system will not be the
naked Fermi surface, but rather a condensate of quark Cooper pairs.

This can be seen intuitively as follows. Consider a system of free
particles.  The Helmholtz free energy is $F= E-\mu N$, where $E$ is
the total energy of the system, $\mu$ is the chemical potential, and
$N$ is the number of particles. The Fermi surface is defined by a
Fermi energy $E_F=\mu$, at which the free energy is minimized, so
adding or subtracting a single particle costs zero free energy.  Now,
suppose a weak attractive interaction is switched on.  BCS showed that
this does not simply lead to a small shift in the Fermi energy.
Rather, it leads to a complete rearrangement of the states near the
Fermi surface. The mechanism is simple: it costs no free energy to
make a pair of particles (or holes), and because of the attractive
interaction it is favorable to do so. Many such pairs will therefore
be created, in all the modes near the Fermi surface, and these pairs,
being bosonic, will form a condensate. The ground state will be a
superposition of states with all numbers of pairs, breaking the
fermion number symmetry. An arbitrarily weak interaction has lead to
spontaneous symmetry breaking.

In condensed matter systems, the BCS mechanism leads to
superconductivity, since it causes Cooper pairing of electrons, which
breaks the electromagnetic gauge symmetry, giving mass to the photon
and producing the Meissner effect (exclusion of magnetic fields from a
superconducting region). It is a rare and delicate state, easily
disrupted by thermal fluctuations, because the dominant interaction
between electrons is the repulsive electrostatic force, and only
in the right kind of crystal are there attractive phonon-mediated
interactions that can overcome it.

In QCD, by contrast, the dominant gauge-boson-mediated interaction
between quarks is itself attractive. This means that we expect quark
pairing to happen whenever the density becomes large enough for the
top of the Fermi sea to consist of quarks rather than nucleons
(``quark matter'').  Since pairs of quarks cannot be color singlets,
the resulting condensate will break the local color symmetry
$SU(3)_{\rm color}$.  We call this ``color superconductivity''.
Note that the quark pairs play the same role here as the Higgs particle
does in the standard model: the color-superconducting phase
can be thought of as the ``Higgsed'' (as opposed to ``confined'')
phase of QCD.

It is important to remember that the breaking of a gauge symmetry
cannot be characterized by a gauge-invariant local order parameter
which vanishes on one side of a phase boundary. The superconducting
phase of QCD can be characterized rigorously only by its global symmetries.
In electromagnetism there is a non-local order parameter, the photon
magnetic mass (Meissner effect), 
% photon gets an electric mass in a plasma
that distinguishes the free phase from the superconducting one.
In QCD there is no free phase: even without pairing the gluons are not
states in the spectrum.  No order parameter distinguishes the Higgsed
phase from a confined phase or a plasma, so we have to look at the
global symmetries.

In this paper we will study QCD with 2+1 quarks.  We take the up and
down quarks to be massless, and study the phase structure of cold,
dense QCD as a function of the strange quark mass. We will start with
the limiting cases of two or three massless quarks.

\section{Two or three massless flavors}

In QCD with two
flavors of massless quarks the Cooper pairs form in the
color ${\bf \bar 3}$ flavor singlet
channel~\cite{mga_Barrois,mga_BarroisPhD,mga_BailinLove,mga_ARW2,mga_RappETC,mga_BergesRajagopal}
\beq
\< q^\al_i q^\be_j \> \sim \ep_{ij}\ep^{\al\be 3}
\eeq
The pattern of symmetry breaking is therefore (with gauge symmetries
in square brackets)
\beq
\ba{rl}
& [SU(3)_{\rm color}]\times [U(1)_Q]
 \times SU(2)_L \times SU(2)_R \\
\to & 
 [SU(2)_{\rm color}]\times [U(1)_{\Qt}]
 \times SU(2)_L \times SU(2)_R \\
\ea
\eeq

\begin{figure}
\begin{center}
\epsfig{file=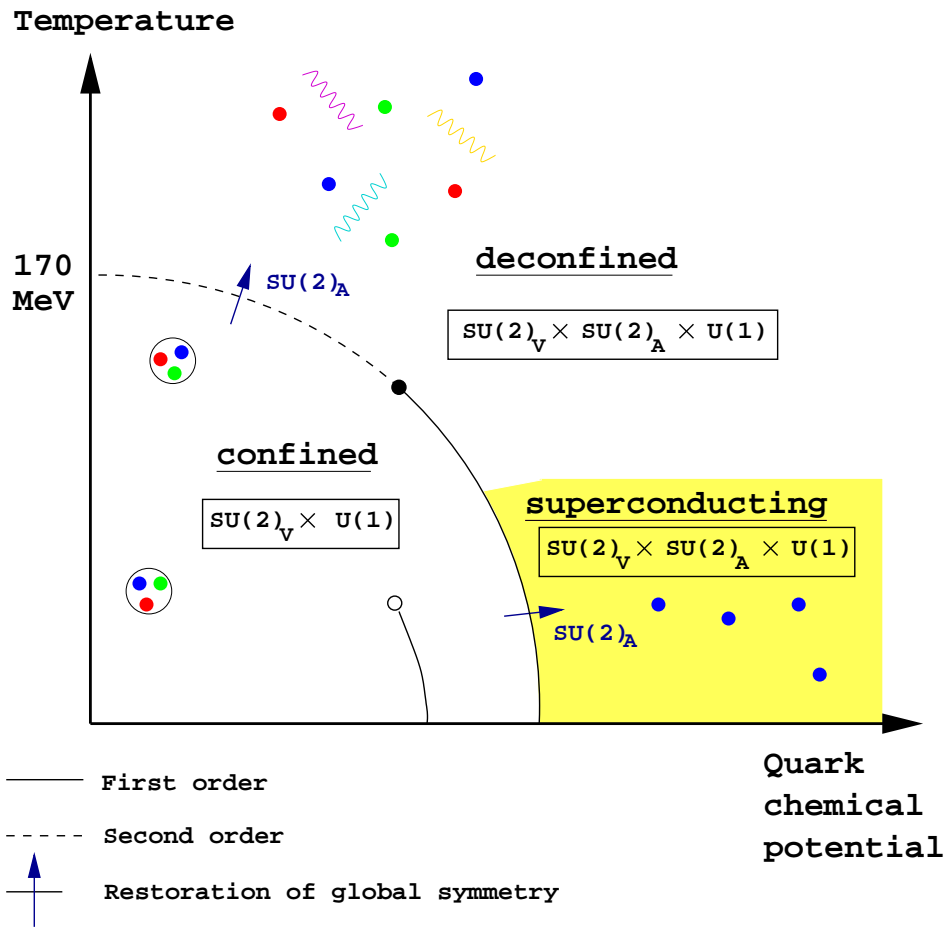,width=5in}
\end{center}
\caption{Two massless flavor phase diagram}
\label{fig:2flav}
\end{figure}

\begin{figure}
\begin{center}
\epsfig{file=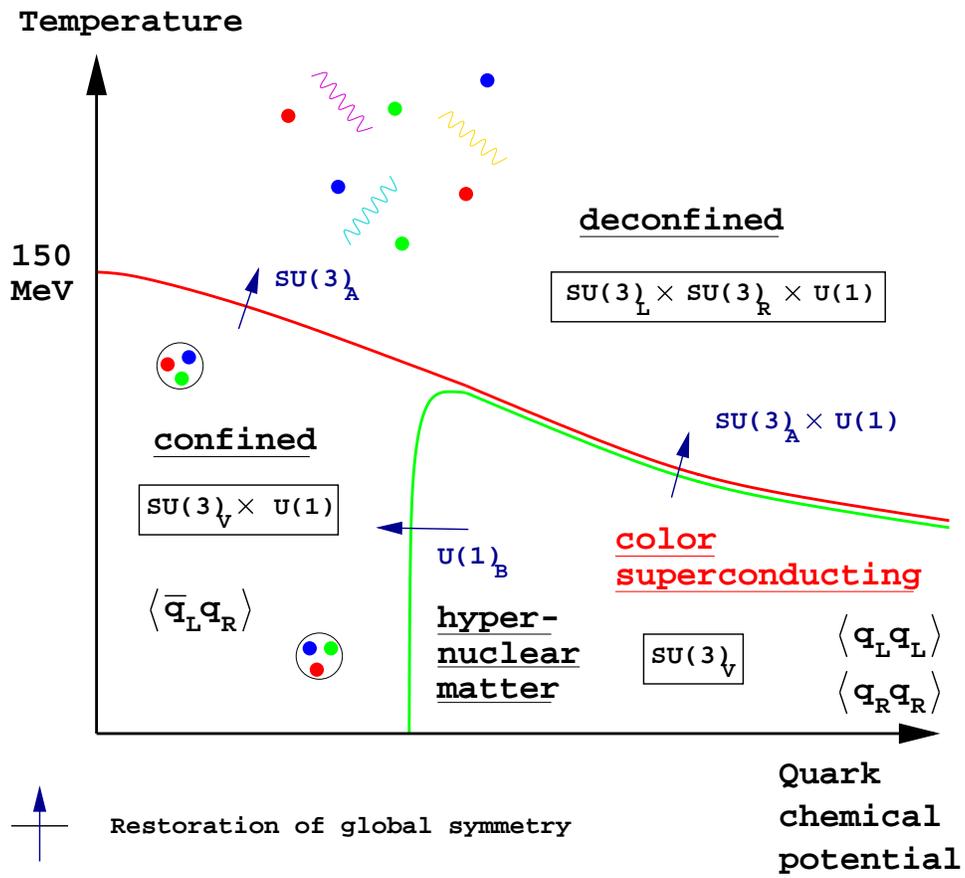,width=5in}
\end{center}
\caption{Three massless flavor phase diagram}
\label{fig:3flav}
\end{figure}

\begin{table}[t]
\caption{Symmetries of phases of QCD. 
With 2 massless flavors, the quark paired phase has the same symmetries
as the QGP, but is different from nuclear matter. With 3 massless flavors, the
quark paired phase has the same symmetries as the hypernuclear matter
phase, but is different from the QGP. 
See Fig.~\ref{fig:2flav}, \ref{fig:3flav}  and
\ref{fig:phasediagram}.
$I_3$ is an isospin generator, $\Y$ is a color generator.}
\vspace{.5pc}
\begin{center}
\bt{cccc}
phase & electromagnetism & chiral symmetry & baryon number \\
\hline
QGP & $Q$ & unbroken & $B$ \\
\hline
\bt{c} 2 flavor\\[-0.5ex] nuclear matter\et 
  & broken & broken & broken \\
\hline
\bt{c} 2 flavor quark\\[-0.5ex]  pairing (2SC) \et
  & $\Qt=Q-\frac{1}{2\sqrt{3}}\Y$ & unbroken   & $\tilde B = \Qt + I_3$ \\
\hline
\bt{c} 3 flavor\\[-0.5ex]  nuclear matter \et 
  & $Q$ & broken & broken \\
\hline
\bt{c} 3 flavor quark\\[-0.5ex]  pairing (CFL) \et
  & $\Qt=Q+\frac{1}{\sqrt{3}}\Y$   & broken  & broken \\
\et
\end{center}
\end{table}
% where $\eta=1/\sqrt{3}$ for CFL, and $\eta=-1/(2\sqrt{3})$ for 2SC.

The resulting condensate gives gaps to quarks with two out of 
the three colors,
and breaks the local color symmetry $SU(3)_{\rm color}$ to an
$SU(2)_{\rm color}$ subgroup.
The Cooper
pairs are $ud-du$ flavor singlets and, in particular, the global
flavor $SU(2)_L \times SU(2)_R$ symmetry is left intact.
There is an unbroken global symmetry which plays the role of baryon
number symmetry, $U(1)_B$.  Thus, no global symmetries are broken and
the only putative Goldstone bosons are those five which become the
longitudinal parts of the five gluons which acquire
masses~\cite{mga_ARW2}. There is also an unbroken gauged symmetry
which plays the role of electromagnetism.  

The third color is prevented from condensing by 'tHooft anomaly matching
conditions~\cite{mga_Sannino}, which require that there be massless fermions
if chiral symmetry is unbroken.

In QCD with three flavors of massless quarks 
the Cooper pairs {\it cannot}
be flavor singlets, and both color and flavor symmetries are
necessarily broken.   The symmetries of the phase which
results have been analyzed in Ref.~\cite{mga_ARW3}
(see also \cite{mga_SS} in which this ordering was studied at zero density).
The attractive channel favored by one-gluon
exchange exhibits ``color-flavor locking''
\beq
\< q^\al_i q^\be_j \> \sim \de^\al_i\de^\be_j + \ka \de^\al_j\de^\be_i
\eeq
This ansatz is only invariant under equal and opposite
color and flavor rotations, and since only the vectorial
part of color is a symmetry of QCD, only the vectorial
part of the flavor group leaves the ansatz invariant.
This pattern of pairing therefore breaks chiral symmetry.
\beq
\ba{rl}
&[SU(3)_{\rm color}]\times [U(1)_Q]
 \times SU(3)_L \times SU(3)_R \times U(1)_B \\
\to & [U(1)_{\Qt}]\times SU(3)_{C+L+R} 
\ea
\eeq

There an unbroken gauged $U(1)$ symmetry (under which
all quarks have integer charges) which plays the role
of electromagnetism.
All nine quarks have a gap.
All eight gluons get a mass. There are nine
massless Nambu Goldstone excitations of the condensate of Cooper pairs
which result from the breaking of the axial $SU(3)_A$ and 
baryon number $U(1)_B$.  
We see that cold
dense quark matter has rather different 
global symmetries for $m_s=0$ than
for $m_s=\infty$.  

\section{2+1 flavor dense matter}

A nonzero strange quark mass explicitly breaks the 
flavor $SU(3)_V$ symmetry. As a consequence, color-flavor locking
with an unbroken global $SU(3)_{{\rm color}+V}$ occurs only for
$m_s\equiv 0$. Instead, for nonzero but sufficiently small strange 
quark mass we expect color-flavor locking  
which leaves a global $SU(2)_{{\rm color}+V}$ group unbroken.
As $m_s$ is increased from zero to infinity, there has to be some 
value $m_s^{\rm unlock}$ at which color and flavor rotations 
are unlocked, and 
the full $SU(2)_L \times SU(2)_R$ symmetry is restored.
It can be argued on general grounds (Sect.~\ref{sec:general}) that
such an simple unlocking phase transition must be first order, although
it is also possible that there may be a crystalline intermediate
phase~\cite{mga_crystal,mga_JBow}.

An analysis of the unlocking transition, using a NJL model with
interaction based on single-gluon exchange~\cite{mga_ABR}, indicates
that for realistic values of the strange quark mass 
chiral symmetry breaking may be present for
densities all the way down to those characteristic of baryonic matter.
This raises the possibility that
quark matter and baryonic matter may be continuously 
connected in nature, as Sch\"afer and Wilczek 
have conjectured 
for QCD with three massless quarks~\cite{mga_SchaeferWilczek}.
The gaps due to pairing at the quark
Fermi surfaces map onto gaps
due to pairing at the baryon Fermi surfaces in 
superfluid baryonic matter consisting of nucleons, 
$\Lambda$'s, $\Sigma$'s, 
and $\Xi$'s. (See Section~\ref{sec:cont}).

Color-flavor 
locking will always occur 
for sufficiently large chemical potential, for 
any nonzero, finite $m_s$ (see Section \ref{sec:highmu}.
As a consequence of color-flavor locking,
chiral symmetry is spontaneously broken even at asymptotically
high densities, in sharp contrast to the well established
restoration of chiral symmetry at high temperature.

\begin{figure}[thb]
\begin{center}
\epsfig{file=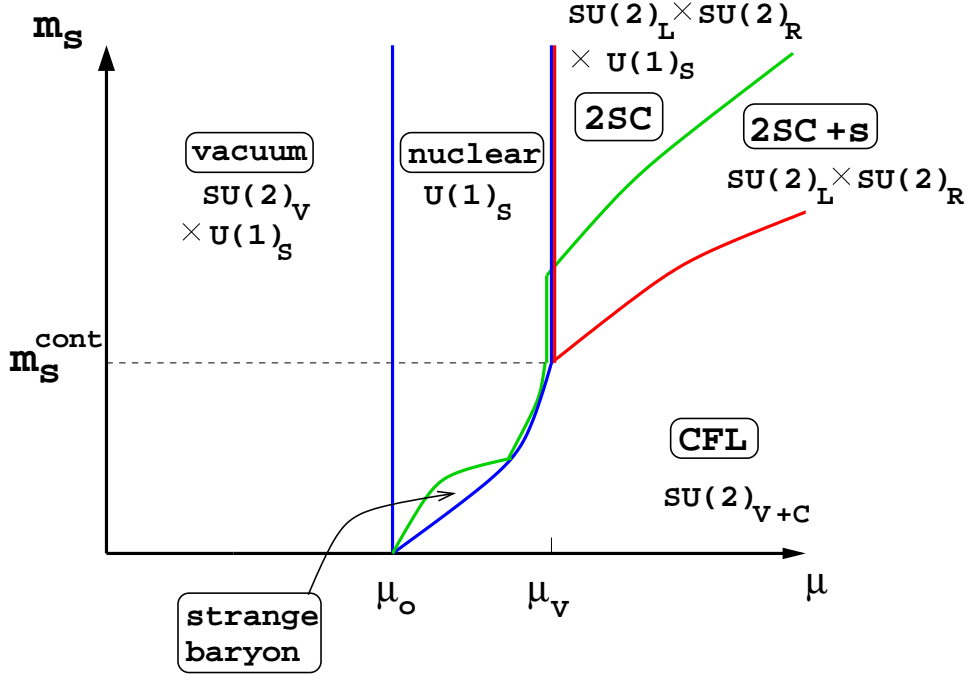,width=5in}
\end{center}
\caption{
 Conjectured phase diagram for QCD with two massless quarks
and a strange quark at zero temperature. 
The global symmetries of each phase are labelled.
The regions of the phase diagram labelled 
2SC, 2SC+s and CFL denote color superconducting quark matter phases.
A detailed explanation is given in the text.
} 
\label{fig:phasediagram}
\end{figure}

Figure \ref{fig:phasediagram}
summarizes our conjecture for the zero temperature
phase diagram of QCD as a function 
of current strange
quark mass $m_s$ and quark number chemical potential $\mu$. 
(we will refer to the $\mu$-dependent constituent
quark mass as $M_s(\mu)$).

Lines in the diagram separate phases which
differ in their global symmetries. 
In each region of the diagram, we list 
the unbroken global symmetries of the corresponding phase.
We characterize the phases using the $SU(2)_L\times SU(2)_R$ 
flavor rotations of the light quarks, and the 
$U(1)_S$ rotations of the strange quarks. The
$U(1)_B$ symmetry associated
with baryon number is a combination of $U(1)_S$, a $U(1)$ subgroup of
isospin, and the gauged $U(1)_{\rm EM}$ of electromagnetism.
Therefore, in our analysis of the global symmetries, 
once we have analyzed isospin and strangeness,
considering baryon number adds nothing new.

In Fig.~\ref{fig:phasediagram} we  
neglect the small $u$ and $d$ current quark masses, since they have
little effect on the condensation of quark Cooper 
pairs~\cite{mga_BergesRajagopal,mga_PisarskiRischke1OPT}. We also ignore the
effects of electromagnetism. 
We assume that wherever a baryon Fermi surface is
present, baryons always pair at zero temperature.
To simplify our analysis, we assume that baryons always
pair in channels which preserve rotational invariance,
breaking internal symmetries such as isospin if necessary. 

We can understand Fig.~\ref{fig:phasediagram} by considering the
phase transitions which occur as $\mu$ is increased 
from 0 to $\infty$ at constant $m_s$
for both large and small $m_s$.

\subsection{Heavy strange quark}
Assume the strange quark is heavy enough that immediately above 
deconfinement $\mu$ is still less than $m_s$, so
there are still no strange quarks present.
For $\mu=0$ the density is
zero; isospin and strangeness are unbroken; Lorentz symmetry is
unbroken; chiral symmetry is broken.

Above a first order transition~\cite{mga_Halasz} at an onset chemical
potential $\mu_{\rm o}\sim 300~\MeV$, one finds nuclear matter.
Lorentz symmetry is broken, leaving only rotational symmetry manifest.
Chiral symmetry is thought to be broken, although the chiral
condensate $\langle \bar q q \rangle$ is expected to be reduced from
its vacuum value.  In the nuclear matter phase there is $pp$ and $nn$
pairing at their Fermi surfaces,
 breaking isospin.  Since there are no strange baryons present,
$U(1)_S$ is unbroken.

When $\mu$ is increased above $\mu_{\rm V}$, a nucleon description is
no longer appropriate. We find the ``2SC'' phase of
color-superconducting quark matter consisting of up and down quarks
only, described in 
Refs.~\cite{mga_Barrois,mga_BarroisPhD,mga_BailinLove,mga_ARW2,mga_RappETC}.  
The
light quarks pair in isosinglet Lorentz singlet channels.
The full chiral flavor symmetry $SU(2)_L \times SU(2)_R$ is unbroken.  
The phase transition at $\mu_V$ is first 
order
\cite{mga_ARW2,mga_RappETC,mga_BergesRajagopal,mga_PisarskiRischke1OPT,mga_BJW2,mga_CarterDiakonov}
and is characterized by a competition between the chiral
$\langle \bar q q\rangle$ condensate and the superconducting
$\langle q q \rangle$ condensate~\cite{mga_BergesRajagopal,mga_CarterDiakonov}.

As the chemical potential is increased further, when $\mu$ exceeds the
constituent strange quark mass $M_s(\mu)$ a strange quark Fermi
surface forms, with a Fermi momentum far below that for the light
quarks.  We denote the resulting phase ``2SC+s''.  Light and strange
quarks do not pair with each other, because their respective Fermi
momenta are so different (see Section \ref{sec:general}).  The strange
Fermi surface is presumably nevertheless unstable. (For more on
single flavor pairing, see Ref.~\cite{mga_IwaIwa}).
The resulting $ss$
condensate is expected to be small~\cite{mga_ABR}, so we
neglect the difference between the 2SC and 2SC+s phases.

Finally, when the chemical potential is high enough that the Fermi
momenta for the strange and light quarks become comparable, we pass through
the first order locking transition
and find the color-flavor locked (CFL) phase.
There is an unbroken global symmetry constructed by locking the
$SU(2)_V$ isospin rotations to an $SU(2)$ subgroup of color. Chiral
symmetry is once again broken.

\subsection{Light strange quark}

We now describe the sequence of phases which arise as $\mu$ is
increased, this time for a value of $\ms$ small enough that strange
baryonic matter forms below the deconfinement density.  At $\mu_{\rm
o}$, one enters the nuclear matter phase, with the familiar $nn$ and
$pp$ pairing that breaks
isospin.  The $\Lambda$, $\Sigma$ and $\Xi$ densities are still zero,
and strangeness is unbroken.  At a somewhat larger chemical potential,
we enter the strange baryonic matter phase, with a strange baryon
Fermi surface (presumably for the $\La$ or $\Si$, self-pairing in a
spin singlet) breaking $U(1)_S$.  This phase is labelled ``strange
baryon'' in Figure \ref{fig:phasediagram}.  The global symmetries
$SU(2)_L\times SU(2)_R$ and $U(1)_S$ are all broken.  As $\mu$ rises,
one may find other onsets at which some of the remaining strange
baryon densities become non-zero. These break no new symmetries, and
so are not shown in the figure.  Note that kaon
condensation~\cite{mga_KaplanNelson} breaks $U(1)_S$, and $SU(2)_V$, and so
by definition it occurs within the "strange baryon"
region of the diagram.  If kaon
condensation is favored, this will tend to enlarge that region.

We can imagine two possibilities for what happens next as $\mu$ increases
further.
(1) Deconfinement: the baryonic Fermi surface is replaced by
$u,d,s$ quark Fermi
surfaces, which are unstable against pairing, and
we enter the CFL phase. Isospin is locked to color and
$SU(2)_{{\rm color}+V}$ is restored, but chiral symmetry remains broken.
(2) No deconfinement: the Fermi momenta of all of the octet
baryons are now similar enough that pairing between baryons with
differing strangeness becomes possible.  At this point,
isospin is restored: the baryons pair in rotationally
invariant, electromagnetically neutral, isosinglets
($p\Xi^-$, $n\Xi^0$, $\Si^+ \Si^-$, $\Si^0\Si^0$, $\La \La$).
The interesting point is that scenario (1) and scenario (2) are 
indistinguishable in their symmetries.
Both look like the ``CFL'' phase of the figure:
$U(1)_S$ and chirality are broken, and there is an
unbroken vector $SU(2)$. This is the ``continuity of quark and hadron matter''
described by Sch\"afer and Wilczek~\cite{mga_SchaeferWilczek}.
We conclude that for low enough strange quark mass, $\ms<\ms^{\rm cont}$, there
may be a region where sufficiently dense baryonic matter has the same
symmetries as quark
matter, and there need not be any 
phase transition between them. In Section 6 we use this observation to
construct a mapping between 
the gaps we have calculated at the Fermi surfaces in
the quark matter phase  and gaps at the baryonic
Fermi surfaces at lower densities.

\section{Model-independent features of the unlocking phase transition}
\label{sec:general}

Assuming that no other intermediate
phases are involved, we can give a model-independent argument that the 
unlocking phase transition between the CFL and 2SC phases
in Figure \ref{fig:phasediagram} must be first order.

A strange quark mass explicitly breaks the
$SU(3)$ flavor symmetry down to the $SU(2)$ involving
only the $u$ and $d$ quarks.
The CFL and 2SC phases are therefore distinguished by whether
the chiral $SU(2)_A$ rotations  are spontaneously
broken.  The unlocking transition
is associated with the vanishing of those diquark
condensates which pair a strange quark with either an up or
a down quark.  We denote the
resulting gaps $\Delta_{us}$ for simplicity.
In the absence of any $\Delta_{us}$ gap,
the only
Cooper pairs are those involving pairs of
light quarks, or pairs of strange quarks.  
The light quark
condensate is unaffected by the strange quarks, and
behaves as in a theory with only two flavors of
quarks. Chiral symmetry is unbroken.  
When $\Delta_{us}\neq 0$,
the interaction between the light quark condensates and
the mixed (light and strange) condensate results in the 
breaking of two-flavor chiral symmetry $SU(2)_A$ via the locking of
$SU(2)_L$  and $SU(2)_R$ 
flavor symmetries to an $SU(2)$ subgroup of color.  This
color-flavor locking mechanism leaves a global
$SU(2)_{{\rm color}+V}$ group unbroken.

\begin{figure}[t]
\begin{center}
\epsfig{file=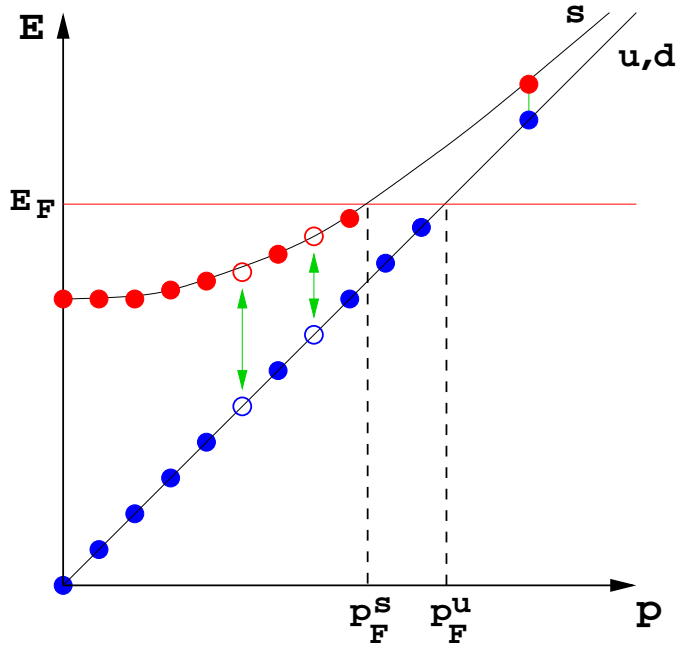,width=3.5in}
\end{center}
\caption{
How the strange quark mass disrupts a $u$-$s$ condensate.  The strange
quark (upper curve) and light quark (straight line) dispersion
relations are shown, with their Fermi seas filled up to the Fermi
energy $E_F$.  The horizontal axis is the magnitude of the
spatial momentum; pairing occurs between particles (or holes)
with the same $p$ and opposite $\vec p$.
For $p<p_F^s$, hole-hole pairing ($\bar s$-$\bar u$) is
possible (two examples are shown).  For $p>p_F^u$, particle-particle
pairing ($s$-$u$) is possible (one example is shown).  Between the
Fermi momenta for the $s$ and $u$ quarks, no such pairing is possible.}
\label{fig:fermi}
\end{figure}

The unlocking transition is a transition 
between a phase with $\Delta_{us}\neq 0$ at  $M_s<M_s^{\rm unlock}$
and a phase with
$\Delta_{us}=0$.  ($M_s(\mu)$
is the constituent strange quark mass at chemical potential $\mu$.)
The BCS mechanism guarantees superconductivity in the 
presence of an arbitrarily weak attractive interaction,
and there is certainly an attraction between $u$ and $s$
quarks (with color ${\bf \bar 3}$) for any $M_s$.
Naively, then, it would seem impossible for
$\Delta_{us}$ to vanish above $M_s^{\rm unlock}$.
The BCS result relies on a singularity
which arises for pairs of fermions with zero total momentum
{\em at} the Fermi surface.
We see from Figure \ref{fig:fermi} that no pairing is possible for quarks
with momenta between the $u$ and $s$ Fermi momenta, and that
at most
one of the quarks in a $u$-$s$ Cooper pair
can be at its respective Fermi surface.
The BCS singularity therefore does not arise if $M_s\neq 0$, 
and a $u$-$s$ condensate is not guaranteed.
A $u$-$s$ condensate involves pairing of quarks with momenta
within about $\Delta_{us}$ of the Fermi surface, 
and we therefore expect that
$\Delta_{us}$ can only be nonzero if the mismatch
between the up and strange Fermi momenta is less than
or of order $\Delta_{us}$:
\begin{equation}
\sqrt{\mu^2-M_u(\mu)^2} - \sqrt{ \mu^2-M_s(\mu)^2} 
 \approx \frac{M_s(\mu)^2-M_u(\mu)^2}{2\mu} 
\lesssim \Delta_{us}\ .
\label{criterion}
\end{equation}
Here $M_s(\mu)$ and $M_u(\mu)$ are the constituent quark masses
in the CFL phase. We neglect $M_u(\mu)$ in the following.
Equation (\ref{criterion}) 
implies that arbitrarily small values of $\Delta_{us}$ are
impossible.
As $m_s$ is increased from zero, $\Delta_{us}$ decreases
until it is comparable to $M_s(\mu)^2/2\mu$. At this point, 
smaller nonzero values of $\Delta_{us}$ are not possible, 
and $\Delta_{us}$  must therefore
vanish discontinuously.  This simple
physical argument leads us to conclude that the
unlocking phase transition at $M_s=M_s^{\rm unlock}$
must be first order.
This was confirmed by explicit calculation
in Ref.~\cite{mga_ABR}, where
the physics of the quark matter phases in Figure
\ref{fig:phasediagram} was analyzed in the mean-field approximation in
a toy model, where the full interactions between quarks were replaced
by a four-fermion interaction with the quantum numbers of single-gluon
exchange.

\section{The 2+1 flavor quark pairing ansatz}
\label{sec:2+1flav}

To analyze the phase diagram one must allow for at least three condensates,
two in quark pair channels and one in the chiral channel,
\begin{equation}
\label{con:conds}
\<q^\al_i C\ga_5 q^\be_j\>\ ,\ \ \ 
\<q^\al_i C\ga_5\ga_4 q^\be_j\>\ ,\ \ \
\< \bar q^{\,i}_\al q^\be_j \>\ ,
\end{equation}
leading to gap parameters
\begin{equation}
\label{con:gaps}
\De^{\al\be}_{ij}\ ,\ \ \ \ \ \ \ \ \ \ 
\ka^{\al\be}_{ij}\ ,\ \ \ \ \ \ \ \ \ \ 
\phi^{i\be}_{\al j}
\end{equation}
Each gap matrix is a 
symmetric $9\times 9$ matrix describing the color (Greek indices)
and flavor (Roman indices) structure.  

In Ref.~\cite{mga_ABR}  we made several simplifying assumptions to
obtain easily soluble gap equations.  These are:
(1) Fix $\phi^{i\be}_{\al j} = \Ms \de^i_3 \de^3_j \de_\alpha^\beta$.
(2) Use the simplest form for $\De^{\al\be}_{ij}$ which
allows an interpolation between the color-flavor locking favored
by single-gluon exchange at $m_s=0$ and the ``2SC'' phase
favored at $m_s\to\infty$ (see \eqn{sol:ansatz}).
This ansatz, which requires
five independent superconducting gap parameters, leads to 
consistent gap equations in the presence of the one-gluon
exchange interaction. 
(3) Neglect $\ka^{\al\be}_{ij}$. This condensate
pairs left-handed and right-handed quarks, and so breaks
chiral symmetry. Its effects
are small~\cite{mga_ABR}.

With these simplifying assumptions, the simplest ansatz that
interpolates between the two flavor case ($\ms=\infty$)
and the three flavor case ($\ms=0$) is
\addtocounter{equation}{3}
\beq\label{sol:ansatz}
%\begin{equation}
\ba{l}
\De^{\al\be}_{ij} =
\left(
\ba{ccccccccc}
b+e & b & c \\
b & b+e & c \\
c & c & d \\
  &   &   & & e \\
  &   &   & e & \\
  &   &   & & & & f\\
  &   &   & & & f &\\
  &   &   & & & & & & f\\
  &   &   & & & & & f &\\
\ea
\right) \\
\hbox{basis vectors:} \\
\ba{rcl@{\,\,}l@{\,\,}l@{\,\,\,\,}
      l@{\,\,}l@{\,\,\,\,} l@{\,\,}l@{\,\,\,\,}l@{\,\,}l}
(\al,i) &=& (1,1),&(2,2),&(3,3),&(1,2),&(2,1),&(1,3),&(3,1),&(2,3),&(3,2) \\
        &=& (r,u),&(g,d),&(b,s),&(r,d),&(g,u),&(r,s),&(b,u),&(g,s),&(b,d)
\ea
\ea
%\label{sol:ansatz}
%\end{equation}
\eeq
where the color indices are $\al,\be$ and the flavor indices are $i,j$.
The strange quark is $i=3$. The rows are labelled by $(\al,i)$ and the
columns by $(\be,j)$.

\begin{table}[hbt]
\def\st{\rule[-1.5ex]{0em}{4ex}} 
\begin{center}
\begin{tabular}{llll}
\hline
\st description & condensate & symmetry  \\
\hline
$\ba{l}\hbox{2SC: 2-flavor} \\ \hbox{\phantom{2SC: }superconductivity}\ea$
 & $\ba{l} c=d=f=0,\\ b=-e \ea$
     & $SU(2)_L\times SU(2)_R$  \\[2ex]
$\ba{l} \hbox{CFL: color-flavor locking}\ea$ &  &   
$SU(2)_{{\rm color}+L+R}$   \\[0.5ex]
$\ba{l}\hbox{CFL: color-flavor locking} \\ \hbox{\phantom{CFL: }with $\ms=0$}\ea$
& $\ba{l} c=b,f=e,\\ d=b+e \ea $ 
     & $SU(3)_{{\rm color}+L+R}$ \\
\hline
\end{tabular}
\end{center}
\caption{
Symmetries of the condensate ansatz \eqn{sol:ansatz} in various
regimes.
}
\label{tab:syms}
\end{table}
The properties of the ansatz are summarized in Table \ref{tab:syms}.
In its general form, this condensate locks color and flavor.
This is because of the condensates $c$ and $f$, referred to 
collectively as $\Delta_{us}$ above,
that
combine a strange quark with a light one.
It is straightforward
to confirm by direct calculation that if either $c$ or $f$ or $b+e$ is
nonzero, then the matrix $\De^{\al\be}_{ij}$ of (\ref{sol:ansatz}) is
not invariant under separate flavor or color rotations but is 
left invariant by simultaneous rotations of $SU(2)_V$ and the $SU(2)$
subgroup of color corresponding to 
the colors 1 and 2.  Thus, color-flavor locking occurs whenever  one
or more of $c$, $f$, or $b+e$ is nonzero.

Although the standard electromagnetic symmetry is broken
in the CFL phase, as are all the color gauge symmetries,
there is a combination of electromagnetic and color
symmetry that is preserved~\cite{mga_ARW3} (see Table \ref{tab:syms}).
Consider
the gauged $U(1)$ under which the charge $\Qt$
of each quark is the sum of its electromagnetic 
charge $(2/3,-1/3,-1/3)$ (depending on the flavor
of the quark) and its color hypercharge $(-2/3,1/3,1/3)$ (depending
on the color of the quark).  It is easy to confirm that the 
sum of the $\Qt$ charges of each pair of quarks corresponding
to a nonzero entry in (\ref{sol:ansatz}) is zero. This modified
electromagnetism is therefore not broken by the condensate
(see Section \ref{sec:flux}).

\section{Quark-hadron continuity}
\label{sec:cont}

As has been emphasized above, for low enough
$\ms$ the CFL phase may consist of hadronic matter at low $\mu$, and
quark matter at high $\mu$. This raises the possibility
\cite{mga_SchaeferWilczek} that properties of sufficiently dense 
hadronic matter could be found by extrapolation from the quark matter
regime where models like the one considered in this paper can be used
as a guide at moderate densities, and where the QCD gauge coupling
becomes small at very high densities.

\begin{table}[htb]
\newlength{\wid}\settowidth{\wid}{XXX}
\def\st{\rule[-1.5ex]{0em}{4ex}} 
\begin{tabular}{lccc|cccc} %@{\protect\phantom{XX}}
\hline
\st Quark & $SU(2)_{{\rm color}+V}$ & $\Qt$ & gap & 
   Hadron & $SU(2)_{V}$   & $Q$ & gap \\
\hline
\multirow{2}{\wid}{$\left(\ba{c} bu\\[1ex] bd \ea\right)$} & 
\multirow{2}{2em}{\bf 2} &
$+1$ &
\multirow{4}{2em}[-1ex]{$ f$} &
\multirow{2}{4em}{$\left(\ba{c} p\\[1ex] n \ea\right)$} & 
\multirow{2}{2em}{\bf 2} &
$+1$ &
\multirow{4}{2em}[-1ex]{$\De^B_4$} \st \\
& & 0 & & & & 0 \st \\
%\cline{1-3}\cline{5-7}
\multirow{2}{\wid}{$\left(\ba{c} gs\\[1ex] rs \ea\right)$} & 
\multirow{2}{2em}{\bf 2} &
0 & &
\multirow{2}{4em}{$\left(\ba{c} \Xi^0\! \\[1ex] \Xi^-\!\! \ea\right)$} & 
\multirow{2}{2em}{\bf 2} &
0 \st \\
& & $-1$ & & & & $-1$ \st \\
\hline
\multirow{3}{\wid}{$\left(\ba{c} ru-gd\\[1ex] gu\\[1ex] rd \ea\right)$} & 
\multirow{3}{2em}{\bf 3} &
0 & 
\multirow{3}{2em}{$ e$}&
\multirow{3}{4em}{$\left(\ba{c} \Si^0 \\[1ex] \Si^+ \\[1ex] \Si^- \ea\right)$} & 
\multirow{3}{2em}{\bf 3} &
0 &
\multirow{3}{2em}{$\De^B_3$}\st \\
& & $+1$ & & & & $+1$ \st \\
& & $-1$ & & & & $-1$ \st \\
\hline
$ru+gd+\xi_- bs$\hspace{-1em} & \hspace{-2em} {\bf 1} & 0 & $\De_-$ & 
  $\La$ \hspace{-1em} & \hspace{-2em} {\bf 1} & 0 & $\De^B_1$ \st \\
\hline
$ru+gd-\xi_+ bs$ & \hspace{-2em}{\bf 1} & 0 & $\De_+$ &
  --- &  \st \\
\hline
\end{tabular}
\[
\ba{rcr@{}l}
\De_\pm &=& \half  & \Bigl( 2b+e+d \pm \sqrt{(2b+e-d)^2+8c^2}\Bigr) \\
\xi_\pm &=& -{1\over 2c} & \Bigl( 2b+e-d  \mp \sqrt{(2b+e-d)^2+8c^2}\Bigr)
\ea
\]
\caption{Comparison of states and gap parameters in high density quark 
and hadronic matter.}
\label{tab:qm}
\end{table}

The most straightforward application of this idea is to relate the
quark/gluon description of the spectrum to the hadron 
description of the spectrum in the CFL 
phase~\cite{mga_SchaeferWilczek}.
As $\mu$ is decreased from the regime in which
a quark/gluon 
description is convenient to one in which a baryonic
description is convenient, there is no change in symmetry 
so there need be no transition: the spectrum of the theory may
change continuously. Under this mapping, 
the massive gluons in the CFL phase
map to the octet of vector
bosons;\footnote{The singlet vector boson in the hadronic phase
does not correspond to a massive gluon in the CFL phase. This
has been discussed in Ref.~\cite{mga_SchaeferWilczek}.}
the Goldstone bosons associated with chiral symmetry breaking
in the CFL phase 
map to the pions;  
and the quarks map onto baryons. Pairing
occurs at the Fermi surfaces, and we therefore expect the gap
parameters in the
various quark channels to map to
the gap parameters due to baryon pairing.

In Table~\ref{tab:qm}
we show how this works for the fermionic states in 2+1 flavor QCD.
There are nine states in the quark matter phase. We show how they
transform under the unbroken ``isospin'' of $SU(2)_{{\rm color}+V}$ and their
charges under the unbroken ``rotated electromagnetism'' generated
by $\Qt$, as described above.
Table~\ref{tab:qm} also shows the baryon octet,
and their transformation properties under the symmetries
of isospin and electromagnetism that are unbroken in sufficiently
dense hadronic matter. Clearly there is a correspondence between
the two sets of particles.

(The one exception is the final isosinglet, discussed at greater
length in Ref.~\cite{mga_ABR}.  In the $\mu\to\infty$ limit, where the
full 3-flavor symmetry is restored, it becomes an $SU(3)$ singlet, so
it is not expected to map to any member of the baryon octet.  The gap
$\De_+$ in this channel is twice as large as the others---it
corresponds to $\De_1$ in Ref.~\cite{mga_ARW3}.)

When we map the quark states onto baryonic states, 
we can predict that the baryonic pairing scheme that will occur
is the one conjectured in Sect.~\ref{sec:int} for sufficiently
dense baryonic matter:
\beq
\ba{ll}
\<p\Xi^-\>,\<\Xi^-p\>,\<n\Xi^0\>,\<\Xi^0n\> 
& \rightarrow \hbox{4 quasiparticles, with gap parameter~~} \De^B_4 \\[0.3ex]
\<\Si^+\Si^-\>,\<\Si^-\Si^+\>,\<\Si^0\Si^0\> 
& \rightarrow\hbox{3 quasiparticles, with gap parameter~~} \De^B_3 \\[0.3ex]
\<\La\La\> & \rightarrow\hbox{1 quasiparticle,\phantom{s} with 
gap parameter~~} \De^B_1
\ea
\eeq
The baryon pairs are rotationally-invariant, $Q$-neutral, $SU(2)_V$
singlets.  It seems reasonable to conclude that as $\mu$ is increased
the baryonic gap parameters $(\De^B_4, \De^B_3, \De^B_1)$ may evolve
continuously to become the quark matter gap parameters $(f, e, \De_-)$.

Is should be born in mind, however,
that the physical quantities are the gaps.
What we have discussed above are the gap parameters, or condensates
(technically, the 1PI two-point functions). The gaps are a function of
these and the particle masses. Since the masses are very different in
the nuclear and quark phases, the gaps may be quite different too,
even if the gap parameters are similar~\cite{mga_gapless}.

\section{Color-Flavor Locking at Asymptotic Densities}
\label{sec:highmu}

\begin{figure}[t]
\begin{center}
\epsfig{file=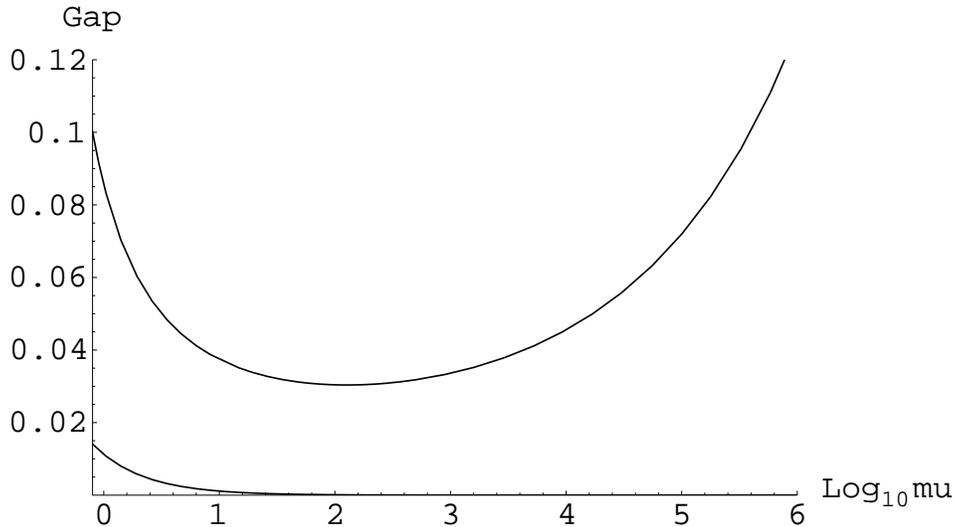,width=5in}
\end{center}
\caption{The upper curve shows the weak-coupling QCD result for the 
superconducting gap $\Delta$ as 
a function of $\log_{10}\mu$, for $\mu$ 
from $0.8$ GeV to $10^6$ GeV. The vertical scale
has been normalized so that $\Delta=0.1$ GeV at $\mu=0.8$ GeV. 
We have taken $g(\mu)$ from the two-loop beta function
for three flavor QCD with $\Lambda_{\rm QCD}=200~\MeV$. Color-flavor
locking occurs whenever $\Delta \gtrsim M_s^2/2\mu$. 
The lower
curve is $M_s^2/2\mu$, taking $M_s=150~\MeV$. 
We conclude that QCD at very high
densities is in the CFL phase.}
\label{fig:son}
\end{figure}

At asymptotically high densities, the QCD coupling $g$ is weak at
the Fermi surface, so diagrammatic methods can be used
\cite{mga_BarroisPhD,mga_Son,mga_weak2flav,mga_weak3flav} to
determine the 
leading behavior of the gap.
\begin{equation}
\Delta \sim C \mu % \left(\frac{1}{g(\mu)}\right)^5 
\frac{1}{g(\mu)^5}
\exp\Bigl(-\frac{3\pi^2}{\sqrt{2}}
\frac{1}{g(\mu)}\Bigr)\ ,
\label{songapequation}
\end{equation}

It is striking that when these calculations are extrapolated to low
density, they give gaps of order $100~\MeV$, in agreement with NJL
calculations~\cite{mga_weak2flav,mga_weak3flav}. 
We have therefore fixed the prefactor $C$ so that
$\Delta=100~\MeV$
at $\mu=800~\MeV$.  We show the result in Figure \ref{fig:son}.
Note that $\Delta$ is plotted versus $\log\mu$; it changes
very slowly.  It decreases by about a factor of three
as $\mu$ is increased to around 100 GeV and then 
begins to rise without bound at even higher densities.
This is basically because $\mu$ rises faster than $\exp(-1/g)$
drops. 
We conclude that independent of any details (like
the precise value of $M_s$, for example) at asymptotically 
high densities $\Delta$ is 
far above $M_s^2/2\mu$.  For any finite value of the strange
quark mass $m_s$, quark matter is in the color-flavor locked
phase, with broken chiral symmetry, at arbitrarily high
densities where the gauge coupling becomes small.

\section{A signature for color superconducting neutron stars}
\label{sec:flux}

\begin{figure}[t]
\begin{center}
\epsfig{file=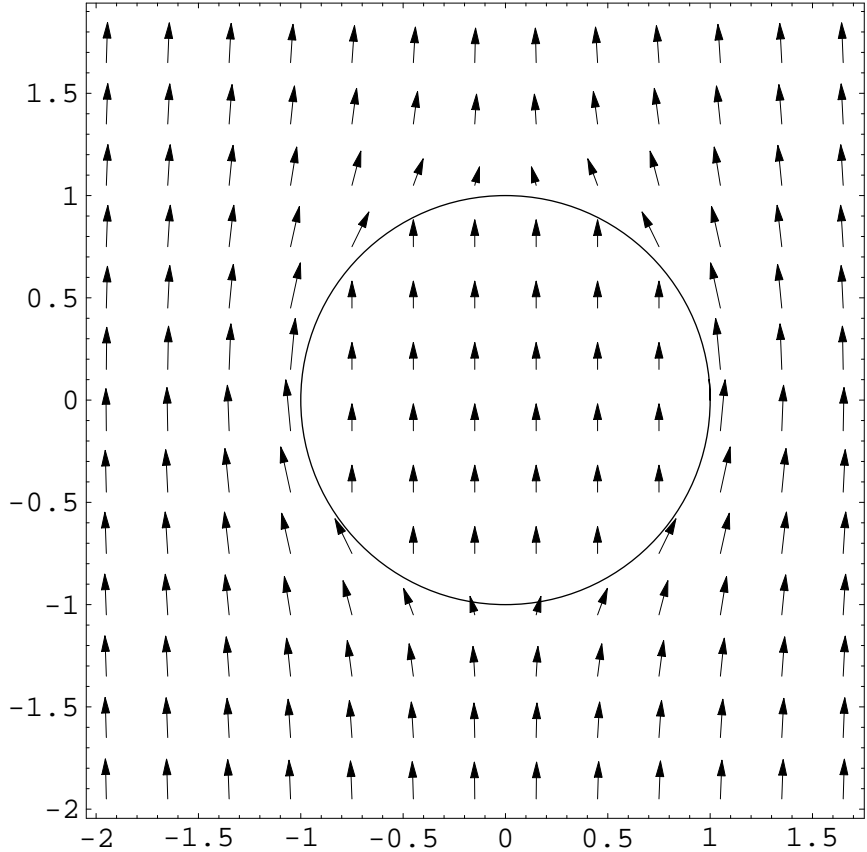,width=4.5in}
\end{center}
\label{fig:flux}
\caption{
Locally unbroken magnetic field inside and outside a sphere
of high-density quark matter, with sharp boundary.}
\end{figure}

The most likely place to find superconducting quark phases in nature
is in the core of neutron stars.  If neutron stars achieve sufficient
central densities that they develop quark matter cores, these cores
must be color superconductors (see Section \ref{sec:int}).  It is
therefore useful to analyze the physical properties of dense quark
matter and try to give observable signatures of the color
superconducting phase, thus allowing astrophysical observation to play
a role.

Though color and ordinary electromagnetism
are broken in a color superconductor, there is a linear combination of
the photon and a gluon that remains massless (Section \ref{sec:2+1flav}). 
Consequently, a color
superconducting region may be penetrated by an external magnetic
field. In Ref.~\cite{mga_flux} it was shown that at most a small fraction of the magnetic field is
expelled, and if the screening distance is the smallest length scale
in the problem there is no expulsion at all.
It was found that color-superconducting regions in a neutron star core
would admit magnetic fields without restricting them to 
quantized flux tubes.  Such magnetic fields  are stable on time scales
longer than the age of the universe, even if
the spin period of the neutron star is changing.

This is interesting because it interferes with the main mechanism
by which the magnetic fields of isolated pulsars are supposed to decay.
This is the dragging of flux tubes by rotational vortices.
As the star spins down, its rotational vortices move outwards,
dragging the flux tubes with them. When the flux tubes reach the
crust, they decay away.
The conductivity in the crust gives a decay time of
$10^6$-$10^{10}$ years~\cite{mga_Ruderman}.

In contrast, we would predict that
as the spin period of the neutron star changes
and the rotational vortices move accordingly, 
there is no change at all in the strength of the 
$\tilde Q$-magnetic field in the core.
The data on isolated pulsars
indicates that the decay time is $10^8$ years or more
\cite{mga_Lorimer}, and is therefore consistent with our hypothesis.

\section{Conclusions}
\label{sec:concl}

We have discussed a conjectured phase diagram for $2+1$ flavor QCD as
a function of $\mu$, the chemical potential for quark number, and
$m_s$, the strange quark mass. We have worked at zero temperature and
ignored electromagnetism and the $u$ and $d$ quark masses throughout. 
The phases may be summarized as follows (see Fig.~\ref{fig:phasediagram}).
\setlength{\itemsep}{-\parsep}
\begin{itemize}
\item There are basically two types of quark matter:
\begin{itemize}
\item CFL: {\em color-flavor locking}\/ quark matter;
  all 3 flavors participate, breaking chiral symmetry.
\item 2SC: {\em two-flavor color superconducting}\/ quark matter;
  $u$-$d$ pairing, chirally symmetric, {\em non}-superfluid.
\end{itemize}
\item The phase transition between CFL and 2SC is 
  {first-order}, assuming no other phase intervenes.
\item If $\ms$ is low enough, then as density rises there
is a transition from baryonic matter to 
chirally broken CFL quark matter. In this case,
\begin{itemize}
\item Quark matter and baryonic matter may be 
  {continuously connected}: symmetries do not require any phase transition
  between them.
\item Chiral symmetry is broken at { all} densities.
\end{itemize}
\item If $\ms$ is high enough, then as density rises 
there is a transition from baryonic matter to 2SC (chirally restored)
quark matter, then to CFL.
\item At sufficiently high density,
chiral symmetry is { \underline{always broken}}.
\end{itemize}

There are many directions in which this work can be developed.  We
have worked at zero temperature, so a natural extension would be to
study the effects of finite temperature. The phase diagram of 
two-flavor QCD as a function of baryon density, temperature 
and quark mass has been explored in Ref.~\cite{mga_BergesRajagopal}.
It would be interesting to perform a similar study of the
3 flavor and 2+1 flavor cases, building on the
suggestions of Pisarski~\cite{mga_Pisarski}.

Within NJL models, one could
study more exotic channels such as
those with $S=1$ and/or $L=1$, or channels that would lead to 
pairing of the strange quarks in 
the 2SC+s phase. A particularly intriguing possibility is that
there is a crystalline phase between the CFL and 2SC phases.
Such phases have been posited in condensed matter contexts
\cite{mga_crystal}, and are currently being studied
in quark matter~\cite{mga_JBow}.

Great progress has been made during the last year in using
weak-coupling methods, appropriate to the limit of asymptotically high
density, to calculate the parameters of the effective theory for the
light degrees of freedom at the Fermi surface~\cite{mga_effth}. But there
remains disagreement over the behavior of the masses of the
pseudo-Goldstone bosons in the large-$\mu$ limit. It would also
be interesting to try to include the effects of instantons in order
to help with the extrapolation to the moderate density regime
that is of physical interest.

It is of great importance to continue to investigate possible
observable consequences of the color-superconducting state.  The
natural arena is the phenomenology of neutron/quark stars, which are
the only naturally occurring example of cold matter at the densities
we have studied. We have already discussed possible effects of the
``rotated'' electromagnetism on magnetic fields. One also expects that
the gaps in the quasiquark spectrum will affect cooling by neutrino
emission, and the shear and bulk viscosities, which play an important
role in the $r$-mode spin-down mechanism~\cite{mga_Madsen}.  Finally,
we should not forget that although current heavy-ion experiments are
oriented primarily towards producing hot low-density fireballs, it is
possible that color-superconducting phases can be made in the cooler
conditions of lower-energy collisions. It would be valuable to
investigate possible signatures: the effect on strangeness production
of the quark-pairing contribution to gluon mass is one obvious
possibility.

\medskip
\begin{center}
{\bf Acknowledgements}
\end{center}
I thank the organizers of the TMU-Yale symposium for making it such a
successful meeting, and J. Berges, K. Rajagopal, and F. Wilczek for
their collaboration on the work I have described here.

%%%%%%%%%%%%%%%%%%%%%%%%%

\end{document}